\documentclass{article}
\usepackage{amsmath}
\usepackage{amsthm}
\usepackage{graphicx}

\newcommand{\omegamax}{\omega_{\rm m}}
\newcommand{\pref}[1]{(\ref{#1})}
\newcommand{\abs}[1]{\left \lvert #1 \right\rvert }

\newtheorem{thm}{Theorem}

\begin{document}

\title{Robust Causality Check for Sampled Scattering Parameters via a Filtered Fourier Transform}
\author{Piero Triverio% <-this % stops a space
\thanks{P.~Triverio is with the Edward S. Rogers Sr. Department of Electrical and Computer Engineering, University of Toronto, Toronto, M5S 3G4 Canada (email: piero.triverio@utoronto.ca).}% <-this % stops a space}
\\ Submitted for publication on July 5th, 2013.
}

\maketitle

\begin{abstract}
We introduce a robust numerical technique to verify the causality of sampled scattering parameters given on a finite bandwidth. The method is based on a filtered Fourier transform and includes a rigorous estimation of the errors caused by missing out-of-band samples. Compared to existing techniques, the method is simpler to implement and provides a useful insight on the time-domain characteristics of the detected violation. Through an applicative example, we shows its usefulness to improve the accuracy and reliability of macromodeling techniques used to convert sampled scattering parameters into models for transient analysis.
\end{abstract}%

\section{Introduction}

Scattering parameters are extensively used in microwave engineering to characterize linear devices such as transmission lines, microstrip filters, antennas, and active devices operating in small-signal conditions. Since all systems in nature are causal, their scattering (S) parameters are expected to satisfy Kramers-Kr\"onig dispersion relations~\cite{jnl-2008-tadvp-causality}, which link the real and imaginary part of every causal frequency response $H(j\omega)$. The time-domain condition for causality is
\begin{equation}
	h(t) = 0 \text{ for } t<0,
	\label{eq:causalitytd}
\end{equation}
where $h(t)$ is the inverse Fourier transform of $H(j\omega)$, commonly called impulse response. 

When S parameters are measured with a vector network analyser or computed with an electromagnetic simulator, their causality may be compromised by several factors including an improper calibration of the instrument, an inaccurate de-embedding of measurement fixtures, noise, and non-causal models for the permittivity and permeability of materials. Non-causal S parameters raise two issues. First, since no device in nature can exhibit such behaviour, they are not physically-consistent and potentially inaccurate. Second, when used within a transient simulation, they can severely impact its accuracy and stability. 
Transient simulations are of paramount importance for the analysis of signal integrity issues in high-speed interconnects, or for the prediction of oscillation buildup in voltage-controlled oscillators. Sampled S parameters are not directly compatible with transient circuit solvers such as SPICE. Therefore, they are frequently converted into an equivalent circuit, called macromodel. Causality violations can severely degrade the accuracy of this macromodeling process, even when performed with well-established techniques like Vector Fitting~\cite{Gus99}, and can even make the final transient simulation diverge~\cite{jnl-2007-tadvp-fundamentals}. The availability of reliable techniques to scan sampled S parameters for causality violations is therefore important to maximize the accuracy and reliability of existing measurement, modeling and simulation processes~\cite{jnl-2008-tadvp-causality}. 

The causality of a given frequency response $H(j\omega)$ can be verified by checking Kramers-Kr\"onig relations~\cite{jnl-2008-tadvp-causality}. Since, in practice, S parameters are available only up to a maximum frequency $\omegamax$, dispersion relations must be truncated introducing a \emph{truncation error} that can compromise the reliability of the causality check. A systematic way to estimate and minimize truncation error in Kramers-Kr\"onig relations was proposed in~\cite{jnl-2008-tadvp-causality} and used to develop a rigorous causality check algorithm for sampled responses. Although very robust, this approach employs a generalized Hilbert transform not easy to implement. 

Alternatively, one can perform an inverse Fourier transform of the samples of $H(j\omega)$ and verify~\pref{eq:causalitytd}. The truncation of the transform to the available bandwidth, however, can heavily bias this test, as we will demonstrate in Sec.~\ref{sec:results}. To the best of our knowledge, there is no published method to systematically account for truncation errors in~\pref{eq:causalitytd}. Previous works either neglected truncation artifacts or used extrapolation to reduce them~\cite{Tes92}. 
In this letter, a rigorous technique to check~\pref{eq:causalitytd} for finite-bandwidth frequency samples is proposed. The method, based on a filtered Fourier transform, features a rigorous estimation of truncation error. Being based on the time-domain condition~\pref{eq:causalitytd}, the method provides valuable information on the time-domain characteristics of the violation not returned by frequency-domain approaches based on dispersion relations. Numerical examples demonstrate the sensitivity of the proposed technique and its application to remove the causality violation in the S parameters of a high-speed connector.

\section{Causality Verification via a Filtered Inverse Fourier Transform}

The simplest way to verify~\pref{eq:causalitytd} is to perform an inverse Fourier transform of the given samples of $H(j\omega)$
\begin{equation}
	h(t) =  \underbrace{\int_{-\omegamax}^{\omegamax} H(j\omega) e^{j\omega t} \frac{d\omega}{2\pi}}_{\displaystyle \hat{h}(t)} +
	\underbrace{ \int_{\abs{\omega} > \omegamax} H(j\omega) e^{j\omega t} \frac{d\omega}{2\pi}}_{\displaystyle e(t)}\,.
	\label{eq:h}
\end{equation}
If samples are available up to a maximum frequency $\omega = \omegamax$, only the first integral can be computed. The obtained impulse response $\hat{h}(t)$ differs from the true one $h(t)$ by a truncation error $e(t)$. Truncation error must be taken into account when verifying~\pref{eq:causalitytd} experimentally, unless $H(j\omega)$ is negligible beyond $\omegamax$. This case rarely happens in practice since S parameters may not even decay to zero as $\omega \to \infty$. As a result, truncation error for S parameters can actually diverge, making a direct check of~\pref{eq:causalitytd} meaningless!

Here, we overcome this issue in a systematic way. Clearly, the influence of missing out-of-band samples must be attenuated. To this end, we apply a low-pass filter $F(j\omega)$ 
to the given scattering response $H(j\omega)$
\begin{equation}
	H_F(j\omega) = F(j\omega) H(j\omega)
	\label{eq:HF}
\end{equation}
and then take the inverse Fourier transform of $H_F(j\omega)$

\begin{equation}
	h_F(t) =  \underbrace{ \int\displaylimits_{-\omegamax}^{\omegamax} H_F(j\omega) e^{j\omega t} \frac{d\omega}{2\pi}}_{\displaystyle \hat{h}_F(t)} +
	\underbrace{ \int\displaylimits_{\abs{\omega} > \omegamax} H_F(j\omega) e^{j\omega t} \frac{d\omega}{2\pi}}_{ \displaystyle e_F(t)}
	\label{eq:hF}
\end{equation}
which returns the impulse response $h_F(t)$ of the filtered response~\pref{eq:HF}. The following theorem shows that, under a suitable condition on the filter, the causality of $h_F(t)$ is \emph{equivalent} to the causality of the given S parameters.%
\begin{thm} \label{thm}%
Let $F(j\omega)$ be minimum phase\footnote{$F(j\omega)$ is minimum phase if both the function itself and its inverse $F^{-1}(j\omega) = 1/F(j\omega)$ are causal~\cite{Rho76}.}. Then, $H(j\omega)$ is causal \emph{if and only if} 
\begin{equation}
	h_F(t) = 0 \text{ for } t<0\,.
	\label{eq:causalityhF}
\end{equation}%
\begin{proof} We proved this Theorem using the bilateral Laplace transform. Because of the limited space, we give here the main proof argument and postpone the fine mathematical details to a full  length paper.
Let $H(j\omega)$ be causal. Since $H_F(j\omega)$ is the product of two causal functions, it is causal and its inverse Fourier transform~\pref{eq:hF} satisfies~\pref{eq:causalityhF}. Conversely, if $h_F(t)$ satisfies~\pref{eq:causalityhF}, its Fourier transform~\pref{eq:HF} is causal. Since $H(j\omega) = F^{-1}(j\omega) H_F(j\omega)$ is the product of two causal functions, it is causal as well.
\end{proof}
\end{thm}%

Theorem~\ref{thm} gives precise conditions under which a filter applied to a frequency response does not alter its causal or non-causal nature. Although filtering is a common practice to reduce the effects of truncation in discrete Fourier transforms, existing filtering windows~\cite{Har78} do not satisfy the minimum phase condition of Theorem~\ref{thm}, and are not suitable for causality verification. All Butterworth and Chebyshev low-pass filters~\cite{Rho76}, instead, satisfy the conditions of Theorem~\ref{thm}.

\subsection{Truncation Error Bound}

Thanks to Theorem~\ref{thm}, we can verify the causality of $H(j\omega)$ through the filtered response~\pref{eq:hF}, using a low-pass filter to control truncation error. We assume that $H(j\omega)$ is bounded by $M$ for $\omega > \omegamax$.

For passive devices, this holds with $M = 1$. For active devices, $M$ can be estimated from the maximum device gain beyond $\omegamax$. Using this information, we bound the truncation error $e_F(t)$ in~\pref{eq:hF} with the chain of inequalities
\begin{equation}
	\abs{e_F(t)} \le \int\displaylimits_{\abs{\omega} > \omegamax} \abs{F(j\omega)H(j\omega)} \frac{d\omega}{2\pi} 
	\le M \int\displaylimits_{\abs{\omega} > \omegamax} \abs{F(j\omega)} \frac{d\omega}{2\pi}
	\label{eq:chain}
\end{equation}
If $F(j\omega)$ is the response of a Butterworth or Chebyshev low-pass filter of order greater than one, the last integral in~\pref{eq:chain} is finite and provides an upper bound $E$ to the truncation error
\begin{equation}
	%\abs{e_F(t)} \le
	E =  M \int_{\abs{\omega} > \omegamax} \abs{F(j\omega)} \frac{d\omega}{2\pi}\,.
	\label{eq:E}
\end{equation}

\subsection{Proposed Causality Check}

From the samples of $H(j\omega)$, we compute the filtered impulse response $\hat{h}_F(t)$ in~\pref{eq:hF} and the bound~\pref{eq:E} for the associated truncation error. The causality check is performed by verifying if $\hat{h}_F(t)$ is below the truncation error threshold $E$ for $t<0$
\begin{equation}
	\abs{\hat{h}_F(t)} \le E \text{ for } t<0\,.
	\label{eq:causalitycheck}
\end{equation}
If~\pref{eq:causalitycheck} is violated for some $t^* < 0$, one can show that~\pref{eq:causalityhF} does not hold for $t = t^*$ and that, by Theorem~\ref{thm}, the given S parameters are not causal. Condition~\pref{eq:causalitycheck} is a robust version of~\pref{eq:causalityhF} which accounts for the finite bandwidth of the given response.

The sensitivity of the check to different types of violations can be controlled through the order and cut-off frequency of the filter. While a low order maximizes sensitivity over the whole samples bandwidth, a high order minimizes the truncation error bound $E$.

\section{Numerical Results}
\label{sec:results}

\subsection{Analytic Example}

\begin{figure}[t]
\includegraphics[width=0.9\columnwidth]{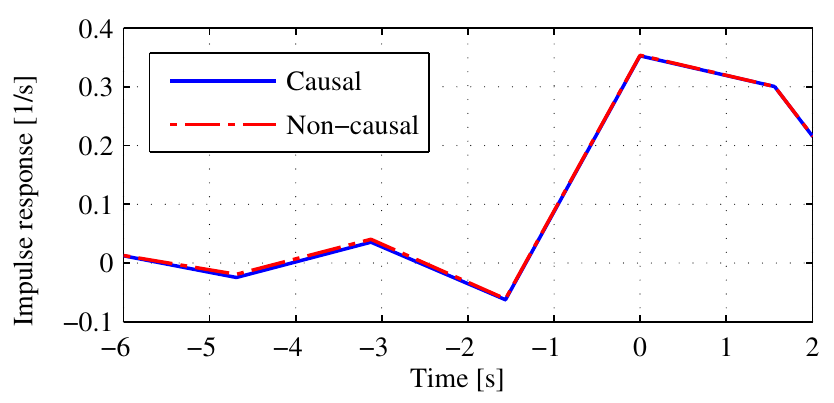}
\caption{Impulse response obtained with IFFT from the causal~\pref{eq:ex_Hc} and non-causal~\pref{eq:ex_Hnc} samples. The difference between the curves is hardly visible.}
\label{fig:ex_ifft}
\includegraphics[width=0.9\columnwidth]{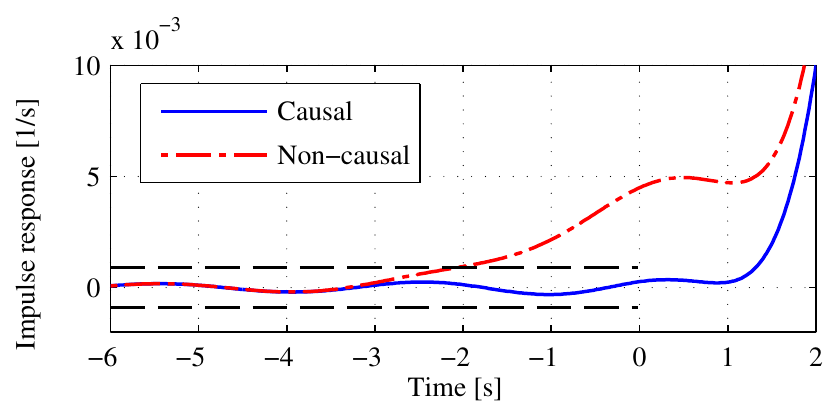}
\caption{Filtered response $\hat{h}_F(t)$ obtained from the causal~\pref{eq:ex_Hc} and non-causal~\pref{eq:ex_Hnc} samples. Dashed lines denote the detection threshold $E$ in~\pref{eq:causalitycheck}.}
\label{fig:ex}
\end{figure}

Through an analytic example, where we can precisely control the presence of causality violations, we demonstrate the sensitivity of the proposed approach and its superior performance compared to a standard inverse Fourier transform. We consider two frequency responses

\begin{align}
	H_c(j\omega) & = \frac{1}{j\omega+1} \,,
	\label{eq:ex_Hc} \\
	H_{nc}(j\omega) & = \frac{1}{j\omega+1} + \frac{0.01}{j\omega+0.5}e^{j\omega 5}\,.
	\label{eq:ex_Hnc}
\end{align}
The first response $H_c(j\omega)$ is causal. The second response $H_{nc}(j\omega)$ is made by the first response perturbed with a very small causality violation. The violation arises because of the anticipation factor $e^{j \omega 5}$. Both frequency responses were sampled at 100 points from DC up to $\omegamax = 2\,{\rm rad/s}$. First, we applied a standard IFFT to the samples obtaining the impulse responses shown in Fig.~\ref{fig:ex_ifft}. Because of truncation errors, both responses apparently violate~\pref{eq:causalitytd}, even though the first one is perfectly causal! The small violation in~\pref{eq:ex_Hnc} is not visible since it is masked by large truncation artifacts.

Next, we applied the proposed methodology using a Chebyshev low-pass filter of order 6, cut-off frequency of 1.4~rad/s, and maximum passband ripple of 3~dB. The filtered impulse responses obtained from the causal and non-causal samples are depicted in Fig.~\ref{fig:ex}. The dashed horizontal lines represent the truncation error bound $E$, and allow for a visual check of~\pref{eq:causalitycheck}. The proposed causality condition is satisfied by the causal samples, but violated by the non-causal samples. The weak violation present in~\pref{eq:ex_Hnc}, undetectable with a standard IFFT, is clearly revealed by the proposed technique.

\subsection{High-Speed Connector}
\label{sec:v}

\begin{figure}[t]
\includegraphics[width=0.9\columnwidth]{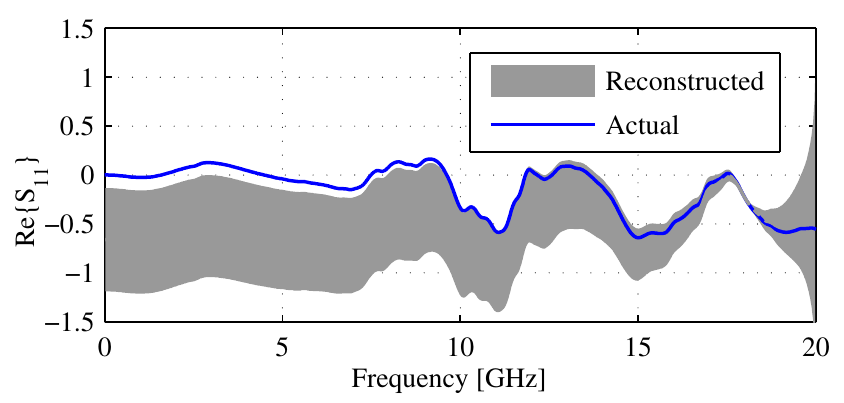}
\caption{Causality check for the S-parameters of the high-speed connector of Sec.~\ref{sec:v}, performed using Kramers-Kr\"onig relations~\cite{jnl-2008-tadvp-causality}: actual real part of the samples (solid blue line) and its reconstruction assuming causality (grey band). Since the two curves are not compatible below 10 GHz, causality violations are present.}
\label{fig:v_gdr}
\includegraphics[width=0.9\columnwidth]{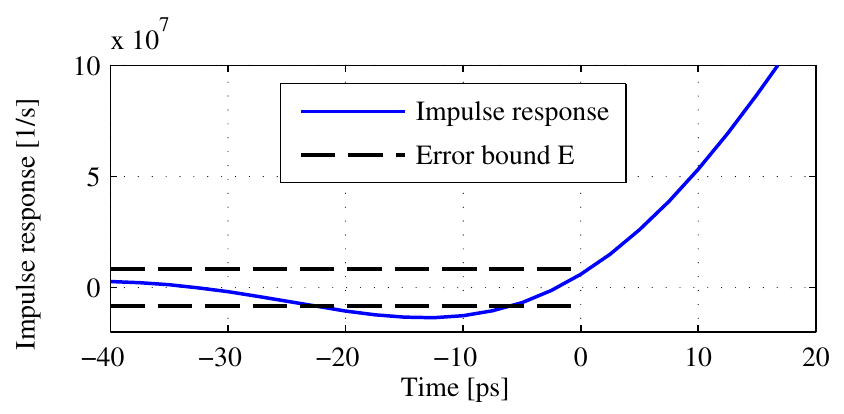}
\caption{High-speed connector of Sec.~\ref{sec:v}: filtered response $\hat{h}_F(t)$ obtained from the samples of $S_{11}$. Dashed lines denote the detection threshold $E$ in~\pref{eq:causalitycheck}.}
\label{fig:v}
\end{figure}

We consider the S parameters of a high-speed connector (courtesy of IBM) which were computed with a field solver up to 20~GHz. We checked the causality of the return loss $S_{11}$ samples using a state-of-the-art technique~\cite{jnl-2008-tadvp-causality} and the proposed method. 

The first method reconstructs the real part of the samples from their imaginary part through Kramers-Kr\"onig relations assuming causality holds. If the reconstructed samples do not match the actual samples, causality is violated. This is indeed the case for the given S parameters, as shown in Fig.~\ref{fig:v_gdr}, where the thickness of the reconstructed curve accounts for truncation errors. Since the two curves are not compatible below 10~GHz, causality is violated. As this method works entirely in the frequency domain, no insight on the time-domain characteristics of the violation is provided.

The proposed technique was applied using a Chebyshev low-pass filter of order 4, cut-off frequency of 6~GHz, and maximum passband ripple of 6~dB. The check took 84~ms and led to the filtered response $\hat{h}_F(t)$ in Fig.~\ref{fig:v}. Causality condition~\pref{eq:causalitycheck} is clearly violated, since $\hat{h}_F(t)$ goes beyond the truncation error bound starting from $t=-22.5\,{\rm ps}$. Although such violation may seem negligible, it actually makes the extraction of an accurate macromodel for transient analysis impossible. Using the robust Vector Fitting algorithm~\cite{Gus99}, we could not obtain a macromodel with an error lower than~73\%. The obtained macromodel had also passivity violations that can make transient simulations diverge~\cite{jnl-2007-tadvp-fundamentals}. As explain in~\cite{jnl-2007-tadvp-fundamentals}, Vector Fitting fails because of the causality violation. 
This was verified by applying a small delay of 22.5~ps to the S parameters. With the delayed samples, Vector Fitting delivers a macromodels with 5.5\% accuracy and no passivity violations. 
This example shows how the proposed method can help in diagnosing and removing causality violations before they compromise the accuracy and stability of modeling and simulation tasks~\cite{jnl-2007-tadvp-fundamentals}. The developed method provides a crucial information on the time extent of the violation which is not immediately available from a frequency-domain analysis like the one shown in Fig.~\ref{fig:v_gdr}.

\section{Conclusion}

We presented a novel methodology to scan sampled scattering parameters for causality violations. The method combines an inverse Fourier transform with low-pass filtering, and rigorously accounts for the uncertainty caused by the finite bandwidth of the samples. Compared to existing techniques based on dispersion relations, the novel method is simpler to implement and provides valuable information on the time-domain characteristics of the violation. %An applicative example shows its usefulness in the modeling  to a high-speed connector shows its  usefulness to improve the modeling of linear components for mixed-signal analyses.

\bibliographystyle{IEEEtran}

\bibliography{IEEEabrv,biblio}

\end{document}